\documentclass[aps,pra,reprint,longbibliography,floats,floatfix]{revtex4-2}

\usepackage{amsmath,amssymb}
\usepackage{graphicx}
\usepackage{booktabs}
\usepackage{bm}
\usepackage{tikz}
\usepackage{quantikz}
\usepackage{hyperref}
\usepackage{microtype}

\hypersetup{colorlinks=true,citecolor=blue,linkcolor=blue,urlcolor=blue}

\newcommand{\cczs}{\mathrm{CCZ}_{6}}
\newcommand{\cczl}{\mathrm{CCZ}_{8}}

\newcommand{\epsHG}{\epsilon_{\mathrm{HG}}}

\def\be{ \begin{equation} }
\def\ee{ \end{equation} }
\def\bse{ \begin{subequations} }
\def\ese{ \end{subequations} }

\begin{document}

\title{\textit{Connectivity Reverses the Performance Ranking of Exact Toffoli Decompositions}}
\title{When More Becomes Less: Topology-Reversed Three-Qubit Gate Performance on IBM Quantum Processors}

\author{Simona Grigorova}
\affiliation{Center for Quantum Technologies, Faculty of Physics, Sofia University St. Kliment Ohridski, 5 James Bourchier Boulevard, 1164 Sofia, Bulgaria}

\author{Stancho G. Stanchev}
\affiliation{Center for Quantum Technologies, Faculty of Physics, Sofia University St. Kliment Ohridski, 5 James Bourchier Boulevard, 1164 Sofia, Bulgaria}

\author{Nikolay V. Vitanov}
\affiliation{Center for Quantum Technologies, Faculty of Physics, Sofia University St. Kliment Ohridski, 5 James Bourchier Boulevard, 1164 Sofia, Bulgaria}

\date{\today}

\begin{abstract}
The exact Toffoli gate admits a six-CX decomposition, denoted by $CCX_6$, that is optimal under unrestricted two-qubit connectivity.
On a linear three-qubit topology, however, $CCX_6$ contains 4 nearest-neighbor CX gates and 2 non-nearest-neighbor CX gates.
By contrast, an alternative exact decomposition, denoted by $CCX_8$, uses only 8 nearest-neighbor CX gates.
Because CCX is locally equivalent to CCZ, we perform the experiments using the corresponding $\cczs$ and $\cczl$ circuits.
We compare these circuits on sampled linear triples of the 156-qubit IBM Quantum Heron processors \texttt{ibm\_fez} and \texttt{ibm\_kingston}.
Under the compilation protocol, the nominal $\cczs$ circuit becomes a twelve-CZ implementation, whereas the linear-nearest-neighbor circuit retains eight native CZ gates. 
Experimentally measured ensemble-feature-selection estimates favor the eight-CZ realization on nearly all retained triples. 
We test the same ordering by preparing a three-qubit hypergraph state, which probes the coherent conditional phase rather than only computational-basis populations. 
The measured hypergraph-state infidelity is lower for the $\cczl$ circuit for most triples on both processors. 
Phase-altered interleaved randomized benchmarking provides a complementary comparison of Clifford surrogates preserving the two compiled entangling structures. 
Within the scope of the tested circuits and phase-sensitive input state, the results demonstrate that hardware connectivity can reverse the operational ranking of exact decompositions: a circuit with more abstract two-qubit gates can yield the better physical implementation.
\end{abstract}

\maketitle

\section{Introduction}

The Toffoli gate is a fundamental ingredient in classical reversible and quantum computation. 
It is used extensively in quantum arithmetic, Boolean-oracle construction, error correction, and fault-tolerant non-Clifford computation \cite{Toffoli1980,Barenco1995}. 
Ripple-carry adders and related arithmetic circuits are naturally expressed in terms of CNOT and Toffoli operations \cite{Cuccaro2004}, while modern fault-tolerant resource estimates for Shor-type factoring count much of the computational workload directly in Toffoli gates \cite{GidneyEkera2021}. 
Multi-controlled NOT and phase operations also provide natural constructions for amplitude-amplification algorithms, and Toffoli operations have been used experimentally in complete Grover-search circuits \cite{Figgatt2017}. 
The gate has a direct role in coherent quantum error correction, where it can implement a syndrome-conditioned corrective operation \cite{Reed2012}. 
At the logical level, the Toffoli gate and the locally equivalent CCZ gate are important non-Clifford resources, motivating specialized low-$T$ constructions and dedicated $|\mathrm{CCZ}\rangle$ magic-state factories \cite{Jones2013,GidneyFowler2019}. 
Consequently, improvements in the physical implementation of Toffoli-class operations can be of great benefit to such applications.

Experimental realizations of Toffoli- and CCZ-class operations span several physical platforms. 
Early demonstrations include a spin-coherence implementation by nuclear magnetic resonance \cite{Cory1998} and a photonic implementation that exploited an auxiliary level of a higher-dimensional information carrier to reduce the required number of two-qubit operations \cite{Lanyon2009}. 
In trapped ions, Monz \textit{et al.} realized a Toffoli gate by temporarily encoding logical information in the collective motion of an ion string \cite{Monz2009}. 
Subsequent trapped-ion experiments incorporated Toffoli operations into Grover search \cite{Figgatt2017}.
Controlled multiqubit entangling operations related to the Toffoli gate have also been demonstrated using optical-tweezer control of trapped ions \cite{Schwerdt2026}. 
Rydberg blockade has enabled a proof-of-principle three-atom Toffoli gate in a neutral-atom array \cite{Levine2019}, complementing theoretical proposals based on direct few-pulse and three-body-resonance mechanisms \cite{Shi2018}.

In superconducting circuits, Toffoli gates have been realized by temporarily accessing higher transmon levels \cite{Fedorov2012,Reed2012}. 
A conceptually different implementation generated a three-qubit CCZ directly through a resonator-induced continuous-variable geometric phase, without decomposition into a sequence of two-qubit gates \cite{Song2017}. 
More recent superconducting implementations include a high-fidelity three-qubit iToffoli gate produced by simultaneous cross-resonance driving \cite{Kim2022} and scalable generalized controlled-phase gates synthesized using auxiliary-level quantum-AND operations \cite{Chu2023}. 
Toffoli protocols have also been proposed for exchange-coupled silicon spin qubits \cite{GullansPetta2019}. 
A recent optical experiment encoded three logical qubits in the polarization and orbital-angular-momentum degrees of freedom of a single photon \cite{Wang2024}, although this realizes a logical three-qubit operation within multiple degrees of freedom of one carrier rather than an entangling gate between three separate physical carriers.

\begin{figure*}[tbph]
\includegraphics[width=0.80\textwidth]{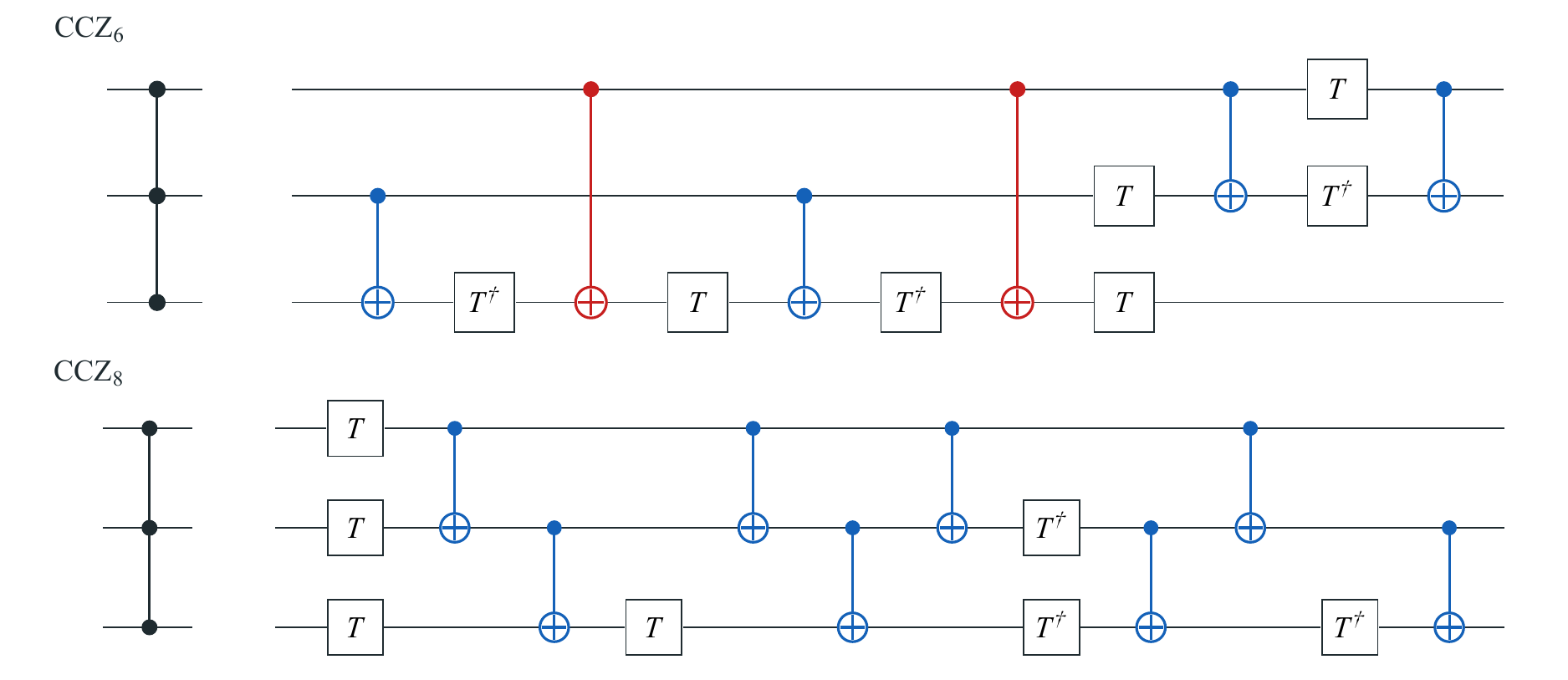}
\caption{Exact $\cczs$ and $\cczl$ decompositions of the CCZ gate. 
$\cczs$ is the standard 6-CX circuit, which contains a non-nearest-neighbor endpoint interaction when embedded on a linear chain, whereas every two-qubit interaction in $\cczl$ acts on adjacent qubits. 
We note that the native two-qubit gates on IBM processors used here are CZ, and hence in the transpilation the CX gates in the circuits in this figure are replaced by CZ sandwiched by two Hadamard gates.
}
\label{fig:T6T8}
\end{figure*}

These implementations illustrate two broad strategies for realizing Toffoli-class operations. 
The first synthesizes the gate entirely within the computational qubit space using sequences of native pairwise entanglers. 
The second reduces the circuit depth by exploiting noncomputational levels, collective motional modes, resonators, multiple degrees of freedom, or native multiqubit interactions. The present work addresses the former setting. 
On the IBM Heron processors considered here, no native Toffoli or CCZ operation is available, and the three-qubit gate is synthesized from one- and two-qubit gates. 
In this setting, hardware connectivity becomes decisive: a decomposition that minimizes the abstract two-qubit-gate count need not minimize the native entangling count, depth, duration, or error after physical placement and routing.

The standard exact Toffoli decomposition, denoted by $\mathrm{CCX}_6$, uses six CX gates and is optimal under unrestricted two-qubit connectivity \cite{ShendeMarkov2009}.
On a nearest-neighbor chain $q_0-q_1-q_2$, however, the endpoint qubits $q_0$ and $q_2$ do not interact directly.
A circuit containing an endpoint operation such as $\mathrm{CX}(q_0,q_2)$ must therefore be routed or resynthesized through the central qubit, potentially increasing the native two-qubit-gate count.
Conversely, exact eight-CX linear-nearest-neighbor (LNN) Toffoli decomposition, denoted by $\mathrm{CCX}_8$, uses only the two physical links of the chain \cite{Duckering2021,Galda2021,CruzMurta2024}.
Since CCX is locally equivalent to CCZ, we work for convenience with the corresponding $\cczs$ and $\cczl$ circuits.
This representation is naturally suited to both the CCZ-specific EFS protocol and the hypergraph-state benchmark used below.
Figure~\ref{fig:T6T8} presents these two CCZ decomposition circuits.
The physically relevant comparison is therefore not simply six versus eight ideal CX gates, but an all-to-all logical gate-count optimum versus a topology-compatible implementation.

This question has important precedents in hardware-aware compilation and synthesis. 
Orchestrated Trios delays three-qubit-gate decomposition until the physical placement is known and selects architecture-tuned templates to reduce two-qubit communication \cite{Duckering2021}. 
QContext additionally exploits the surrounding circuit context when choosing a decomposition \cite{QContext}, while connectivity-aware synthesis has produced families of exact eight-entangler LNN Toffoli circuits \cite{Galda2021,CruzMurta2024}. 
Native-gate and pulse-level constructions provide further routes to reducing the physical cost \cite{Satoh2022,AbuGhanem2025}. 
Context-sensitive compilers may also replace an exact Toffoli gate with a cheaper relative-phase or restricted-subspace realization when equivalence of the complete surrounding circuit is preserved \cite{QContext}. 
These approaches demonstrate that the appropriate decomposition depends on the native gate set, connectivity graph, physical placement, and circuit context rather than on the abstract entangling count alone.

Here we isolate the effect of connectivity by comparing two fixed, exact CCZ decompositions on the 156-qubit IBM Quantum Heron processors \texttt{ibm\_fez} and \texttt{ibm\_kingston} \cite{IBMHeron}. 
For every sampled linear physical triple, the two templates are assigned to the same qubits and compiled under the same constraints. 
Under the fixed compilation protocol, the endpoint interaction in $\cczs$ requires routing, and the nominal six-CX template becomes a twelve-CZ physical implementation. 
By contrast, every interaction in $\cczl$ already respects the physical chain, so the template retains eight native CZ gates. 
The resulting comparison is therefore between two circuits implementing the same three-qubit unitary but using different topology-induced physical resources.

We examine the performance ordering at three complementary levels. 
\textit{First}, the ensemble feature selection (EFS) protocol combines 24 experimentally measured circuit features using the independently determined estimator of Ref.~\cite{StanchevEFS2026}. 
This provides an experimentally based estimate of the CCZ process infidelity. 
\textit{Second}, we prepare the three-qubit CCZ hypergraph state. 
Because an ideal CCZ leaves all computational-basis populations unchanged and modifies only the phase of the $|111\rangle$ component, this experiment probes the defining coherent conditional phase that a computational-basis truth table would miss. 
\textit{Third}, phase-altered interleaved randomized benchmarking (PA-IRB) compares Clifford surrogates that preserve the physical entangling structures of the two compiled decompositions \cite{GrigorovaPAIRB2026}. 
%
All three comparisons consistently favor the topology-compatible $\cczl$ realization at the aggregate level,
%
and demonstrate the central result: minimizing the abstract two-qubit-gate count can select a decomposition whose routed physical realization has a larger native entangling count and poorer experimentally measured performance than a logically larger but topology-compatible decomposition.

\section{Two exact decompositions on a linear topology}
\label{sec:decompositions}

The Toffoli and CCZ gates are locally equivalent,
\begin{equation}
U_{\mathrm{CCX}}=H_t U_{\mathrm{CCZ}} H_t,
\end{equation}
where $t$ is the target qubit. The CCZ unitary acts as
\begin{equation}
U_{\mathrm{CCZ}}\lvert x_0x_1x_2\rangle
=(-1)^{x_0x_1x_2}\lvert x_0x_1x_2\rangle.
\end{equation}
Both implement the same three-qubit unitary up to the identical local Hadamards used to convert between CCX and CCZ. 
The hardware benchmark is performed in the CCZ representation; it therefore tests the conditional-phase implementation directly, rather than the target-flip truth table of CCX.

Figure~\ref{fig:T6T8} summarizes the hardware distinction. The $\cczs$ optimum assumes the complete interaction graph of three qubits. 
When it is assigned to a physical path, its endpoint interaction must be synthesized through the central qubit. 
In the compilation used for the present experiments (see Appendix~\ref{app:triples}), this produces a twelve-CCZ executable circuit. 
The LNN decomposition uses only nearest-neighbor interactions and produces an eight-CZ executable circuit. 
Thus, the number of CZ operations for the gates at hand is 
$N_{\mathrm{CZ}}(\cczs)=12$ and
$N_{\mathrm{CZ}}(\cczl)=8$
for the fixed compilation protocol considered here. 
Exact eight-entangler decompositions of the three-qubit Toffoli/CCZ on a chain are well established in topology-aware and connectivity-specific synthesis \cite{Duckering2021,Galda2021,CruzMurta2024}; the important point for the present experiment is that the same fixed input templates are used on every physical triple and are verified to implement the same three-qubit unitary before execution.




We study connected physical triples $q_L-q_M-q_R$ on \texttt{ibm\_fez} and \texttt{ibm\_kingston}. Heron processors contain 156 qubits and use calibrated CZ interactions on their coupling graph \cite{IBMHeron,IBMBackend}. One group of 40 mutually disjoint triples was sampled on each processor.
For every triple, the two decompositions were assigned to the same physical qubits and compiled with a fixed layout and common optimization settings. 
Global three-qubit resynthesis was disabled so that the identity of the two input templates was preserved. 
The resulting circuits were checked for unitary equivalence before execution. 

\section{Experimentally measured ensemble feature selection benchmark}

\begin{figure}[t]
\centering
\includegraphics[
  width=0.8\columnwidth
]{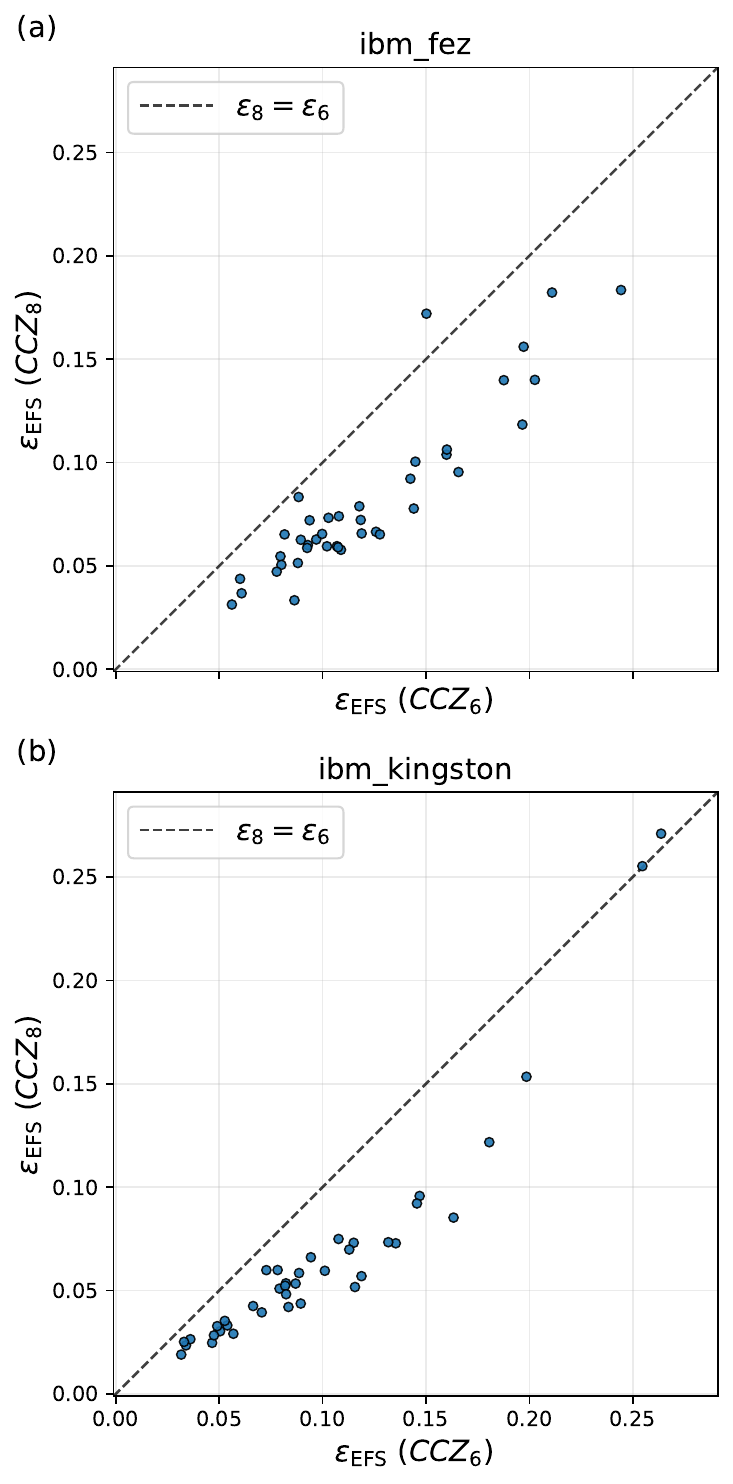}
\caption{EFS estimates of the infidelity of $\cczs$ and
$\cczl$ on (a) \texttt{ibm\_fez} and (b) \texttt{ibm\_kingston}.
Each point compares the two implementations on the same physical
triple. The dashed line denotes equal  infidelity; points below
the line correspond to lower  infidelity for $\cczl$.}
\label{fig:efs}
\end{figure}

We evaluate $\cczs$ and $\cczl$ on \texttt{ibm\_fez} and \texttt{ibm\_kingston} using the Ensemble Feature Selection (EFS) method, developed specifically for the non-Clifford CCZ gate \cite{StanchevEFS2026}.
EFS exploits the linear dependence of process infidelity on experimentally accessible survival probabilities.
A large pool of candidate features is constructed from experimentally executable probe circuits, with each feature defined relative to a corresponding identity circuit.
A physically motivated ensemble of noisy channels is then used offline to select a compact subset of informative features.
The weights combining the selected features into a linear infidelity estimator are learned by ridge regression.
Once the features and weights are fixed, the CCZ infidelity is estimated directly from a small set of experimentally measured circuits, without requiring the target gate to be Clifford or performing full process tomography.
Here we use the 24 features and corresponding weights determined in Ref.~\cite{StanchevEFS2026}.

To place the EFS results on the average-gate-infidelity scale used by PA-IRB, we convert the process-infidelity estimates using Nielsen's relation \cite{nielsen2002} for a three-qubit gate,
\be
\epsilon_{\mathrm{EFS}}
=
\frac{d}{d+1}\,\epsilon_{\mathrm{EFS}}^{\mathrm{pro}}
=
\frac{8}{9}\,\epsilon_{\mathrm{EFS}}^{\mathrm{pro}},
\quad d=2^n=8.
\ee
The measurements were performed with 2000 shots per circuit. 
In the previous EFS study \cite{StanchevEFS2026}, all 204 valid triples were divided into seven groups of disjoint triples for parallel execution. 
Here, we measured in parallel only the first group, comprising 40 triples on \texttt{ibm\_fez} and 39 triples on \texttt{ibm\_kingston}.

\begin{figure}[t]
\centering
\includegraphics[width=0.85\columnwidth]{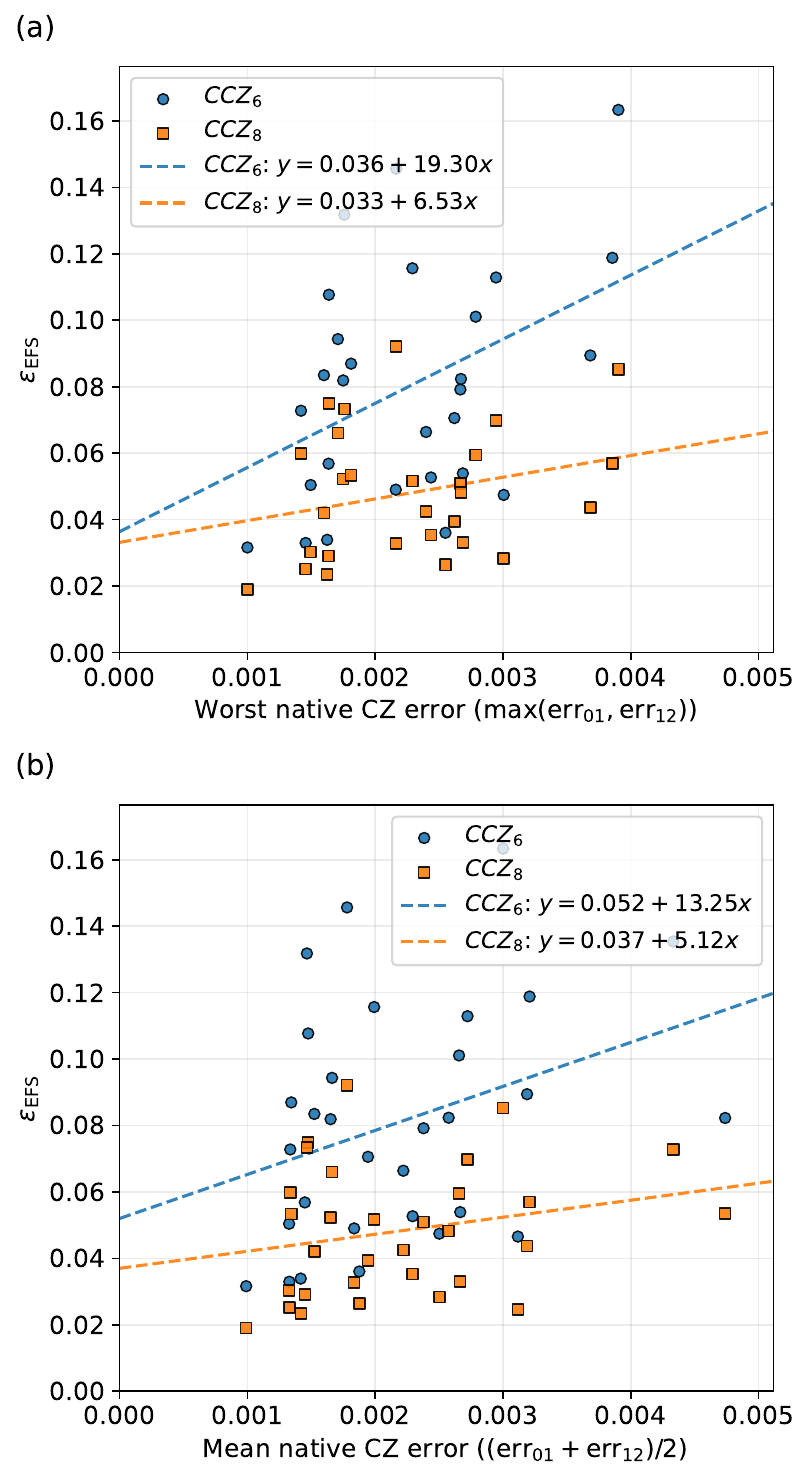}
\caption{EFS errors of the two circuits versus (a) the worse and (b) the mean of the two native CZ errors on \texttt{ibm\_kingston}.
The displayed linear fits of the form $y=a+bx$ are included as exploratory baselines and are not interpreted as physical scaling laws.
}
\label{fig:weakest}
\end{figure}

The sets of physical triples used in the EFS and hypergraph-state experiments in the next section are not identical because the two measurements were performed in separate hardware sessions several days apart. 
Before each experiment, the triples were selected using the calibration data available at that time, excluding placements containing poorly performing qubits or couplers. 
Since the processor calibrations changed between the two sessions, the selected samples contain 40 and 39 triples
for the EFS measurements on \texttt{ibm\_fez} and \texttt{ibm\_kingston}, respectively, and 41 and 38 triples for the
corresponding hypergraph-state measurements. 
All comparisons remain paired within each experiment: $\cczs$ and $\cczl$ are evaluated on the
same physical triples under the same calibration conditions. 
The physical triples used in the EFS and hypergraph-state measurements are listed in Appendix~\ref{app:triples}, in Tables~\ref{tab:triples-fez} and \ref{tab:triples-kingston}.

Figure~\ref{fig:efs} compares the two experimentally measured EFS estimates pairwise. On both processors, nearly all points lie below the equality line. 
On \texttt{ibm\_fez} and \texttt{ibm\_kingston}, $\cczl$ performs
better on 39 of 40, and 37 of 39 triples, respectively. 
The topology-compatible circuit $\cczl$ is therefore favored by the EFS estimator despite its larger abstract CZ count.
The corresponding median reductions in infidelity are $0.040$ and $0.030$, with
95\% bootstrap confidence intervals of $0.033-0.048$ and
$0.022-0.041$.
The measured infidelity estimates span a range comparable to the $0.02-0.20$ range.
The observed median improvement of $\cczl$ over $\cczs$ is therefore clearly resolvable on the accuracy scale established for the
EFS estimator. 


Figure \ref{fig:weakest} plots the EFS data from the two architectures $\cczs$ and $\cczl$ as a function of the worse (top) and the mean (bottom) of the two native CZ infidelities in each triple (in each triple $0-1-2$ the native CZ gates are $0-1$ and $1-2$).
We have plotted the linear fits to each data set, i.e. $\cczs$ and $\cczl$.
As expected from the preceding figure, the $\cczs$ fit lies considerably above the one for the native $\cczl$ architecture, which indicates worse error scaling.

The $\cczs$ line has a slope of 19.3 vs the worse native CZ error and 13.25 vs the average CZ error
The $\cczl$ lines have corresponding slopes of 6.53 and 5.12
This suggests that the routed $\cczs$ realization is more sensitive to the weakest physical link.
The difference in the slopes between $\cczs$ and $\cczl$ in each panel is far greater than $\frac{12}8$, which suggests that the circuit errors grow faster than linearly with the native CZ error --- in fact the slope ratio is closer to a quadratic dependence, $(\frac{12}8)^2 = \frac94$.
However, the dispersion of the data is far too large to claim this feature with certainty. 
It is worth noting that all fits cross the vertical axis at about 0.03-0.05, which looks like a legitimate measure for the SPAM errors, single-qubit gates, shot noise and other errors which do not depend on the CZ count.

\section{Phase-sensitive hypergraph-state benchmark}

A classical truth table is insufficient for CCZ because the ideal gate leaves every computational-basis population unchanged. We therefore probe the coherent phase by preparing the three-qubit hypergraph state \cite{Rossi2013}
\be
\lvert H_3\rangle
=U_{\mathrm{CCZ}}\lvert +++\rangle 
=\frac{1}{\sqrt{8}}\Bigg(
\sum_{x\neq 111}\lvert x\rangle-\lvert111\rangle
\Bigg).
\label{eq:hypergraph}
\ee
The minus sign of the $\lvert111\rangle$ amplitude is the defining nonlinear phase. 
For each physical triple and each circuit, the same state-preparation, measurement, shot budget, and data-processing procedure were used. 
We define the measured hypergraph-state infidelity as
\begin{equation}
\epsHG=1-\langle H_3\rvert\rho_{\mathrm{out}}\lvert H_3\rangle.
\label{eq:hginf}
\end{equation}
Equivalently, with $\Pi_{H_3}=\lvert H_3\rangle\!\langle H_3\rvert$, the fidelity can be written in the three-qubit Pauli basis as
\begin{equation}
F_{\mathrm{HG}}=1-\epsHG
=\frac{1}{8}\sum_{P\in\{I,X,Y,Z\}^{\otimes 3}}
 c_P\,\langle P\rangle,
\label{eq:hgpauli}
\end{equation}
with $c_P=\langle H_3\rvert P\lvert H_3\rangle$.
This identity permits the benchmark to be evaluated either through a reconstructed density matrix or by direct evaluation of the required Pauli expectation values. 
What matters for the paired comparison is that the same estimator, measurement settings, shot budget, and mitigation/data-processing procedure are used for the two circuits on a given physical triple. 

The hypergraph-state measurements were performed in all 27 combinations
of the $X$, $Y$, and $Z$ measurement bases, using 1000 shots per basis.
The density matrices were obtained by linear reconstruction from the
measured Pauli expectation values. Readout mitigation was applied, and
the reported results were calculated from the mitigated probabilities.

Restricting the comparison to the range in which both hypergraph-state
infidelities do not exceed $0.4$ retains 35 of 41 triples on
\texttt{ibm\_fez} and all 38 triples on \texttt{ibm\_kingston}. On
\texttt{ibm\_fez}, $\cczl$ performs better on all 35 retained triples,
with a median reduction in hypergraph-state infidelity of $0.042$ and a
95\% bootstrap confidence interval of $0.022$--$0.050$. On
\texttt{ibm\_kingston}, $\cczl$ performs better on 33 of 38 triples,
with a median reduction of $0.027$ and a 95\% bootstrap confidence
interval of $0.019$--$0.031$. A paired Wilcoxon signed-rank test
confirms the systematic advantage of $\cczl$, yielding
$p=2.9\times10^{-11}$ for \texttt{ibm\_fez} and
$p=1.9\times10^{-6}$ for \texttt{ibm\_kingston}.

\begin{figure}[tbph]
\centering
\includegraphics[width=0.8\columnwidth]{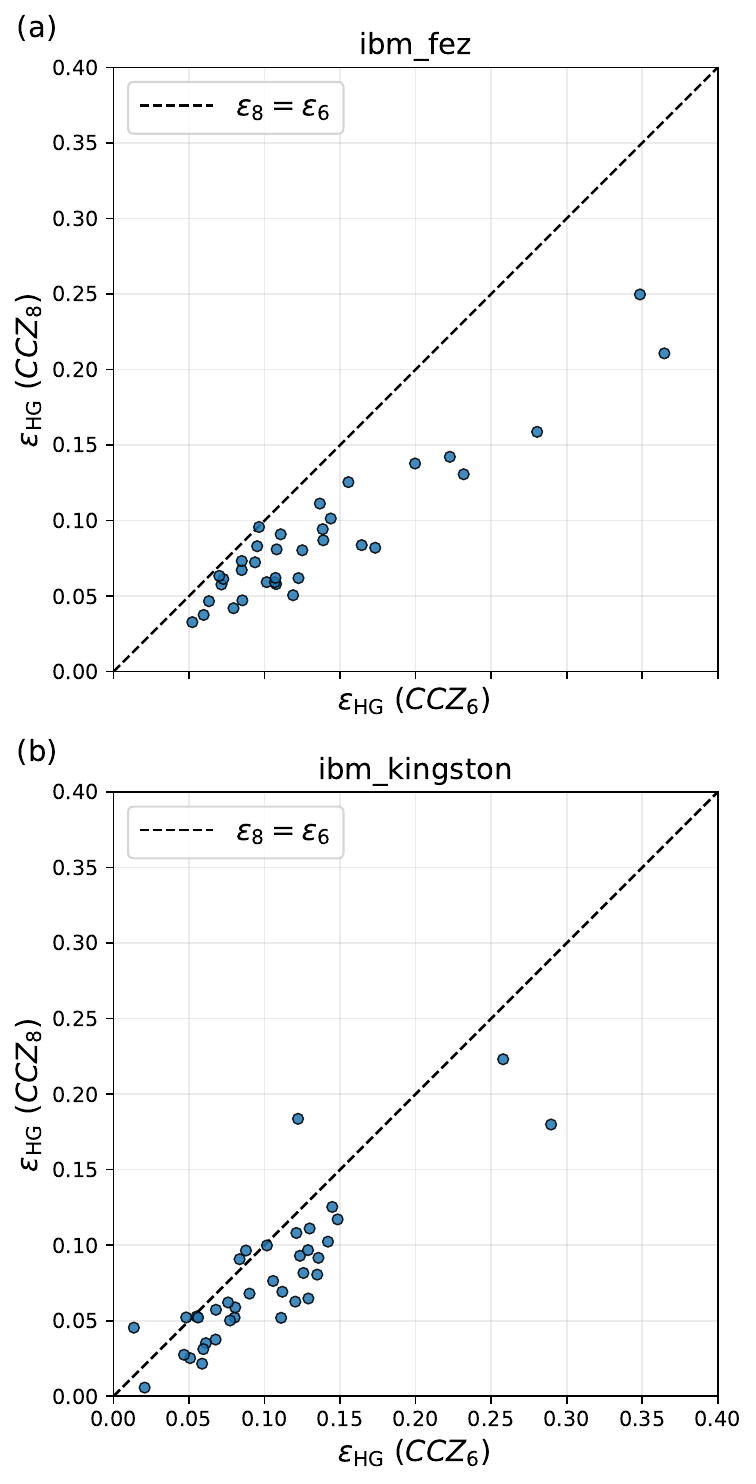}
\caption{Measured hypergraph-state infidelity of the $\cczs$ and $\cczl$ circuits on (a) \texttt{ibm\_fez} and (b) \texttt{ibm\_kingston}. Each point compares the two circuits on the same physical triple. Points below the dashed equality line favor the topology-compatible implementation $\cczl$. 
}
\label{fig:hg}
\end{figure}

Figure~\ref{fig:hg} contains the central direct hardware result. On both processors, most points lie below the equality line, showing that the $\cczl$ implementation produces the target state with lower infidelity than the routed $\cczs$ implementation on most sampled placements. 
The combined mean fidelity rises from approximately $88.2\%$ for $\cczs$ to $91.7\%$ for $\cczl$, as summarized in Table~\ref{tab:mean-performance}. 
This behavior is consistent with the reduced native entangling count, but as for the EFS measurements, it is not a tautological consequence of it: coherent errors, link asymmetry, duration, and one-qubit compilation can in principle reverse the ranking. 

\section{Phase-Altered interleaved randomized benchmarking of the two decompositions}
\label{sec:pairb}

The hypergraph-state experiment probes the ability of the two CCZ
realizations to produce the defining coherent conditional phase on a
particular input state. As a complementary \textit{surrogate-based decay diagnostic}, we
compare the two compiled decompositions using phase-altered interleaved
randomized benchmarking (PA-IRB) \cite{GrigorovaPAIRB2026}. PA-IRB was
introduced for compiled non-Clifford operations whose non-Clifford phase
components can be separated from the Clifford part of the physical
implementation. Rather than interleaving the non-Clifford operation
directly, the method constructs Clifford interleaving gates by altering
the non-Clifford phase rotations while retaining the physical structure
of the compiled circuit.

For the present experiment, PA-IRB is applied separately to the $\cczs$ and $\cczl$ decompositions after compilation to the native basis of \texttt{ibm\_kingston}. 
The nominal $\cczs$ decomposition becomes the routed twelve-CZ implementation discussed in Sec.~\ref{sec:decompositions}, whereas the
LNN $\cczl$ decomposition remains an eight-CZ implementation. 
For each compiled circuit we construct two Clifford surrogates, denoted by $G_{6,s}, G_{6,d}, G_{8,s}, G_{8,d}$,
where $s$ and $d$ indicate the phase-stripped and phase-dressed
constructions, respectively.

The non-Clifford components of the native circuits are the
$R_Z(\pm\pi/4)$ rotations originating from the $T$ and $T^\dagger$
operations. In the phase-stripped construction these quarter-turn
rotations are removed while the remaining native single-qubit gates,
CZ gates, and routing structure are left unchanged. In the
phase-dressed construction an additional rotation of the same sign is
associated with every $R_Z(\pm\pi/4)$ rotation, so that the resulting
pair is equivalent to the Clifford rotation
$R_Z(\pm\pi/2)$. The complete stripped and dressed circuits were
verified numerically to be Clifford operations before being used as
IRB interleaving elements. Thus, for each logical decomposition the
two PA-IRB variants retain the corresponding compiled entangling
skeleton while providing Clifford circuits admissible to the standard
IRB construction.

For each physical triple we first perform reference randomized
benchmarking using random three-qubit Clifford sequences of lengths
$m\in\{1,2,4,8,16,32\}$.
For every sequence length, $K=15$ independently generated Clifford
sequences are used, with 1000 measurement shots per circuit. Four
corresponding interleaved experiments are then performed using
$G_{6,s}$, $G_{6,d}$, $G_{8,s}$, and $G_{8,d}$. The same physical
three-qubit placements are used throughout the reference and
interleaved experiments. Disjoint triples are embedded in parallel
within each submitted circuit, so that each circuit realization probes
the selected placements under a common experimental execution.

\begin{figure}[t]
\centering
\includegraphics[width=0.8\columnwidth]
{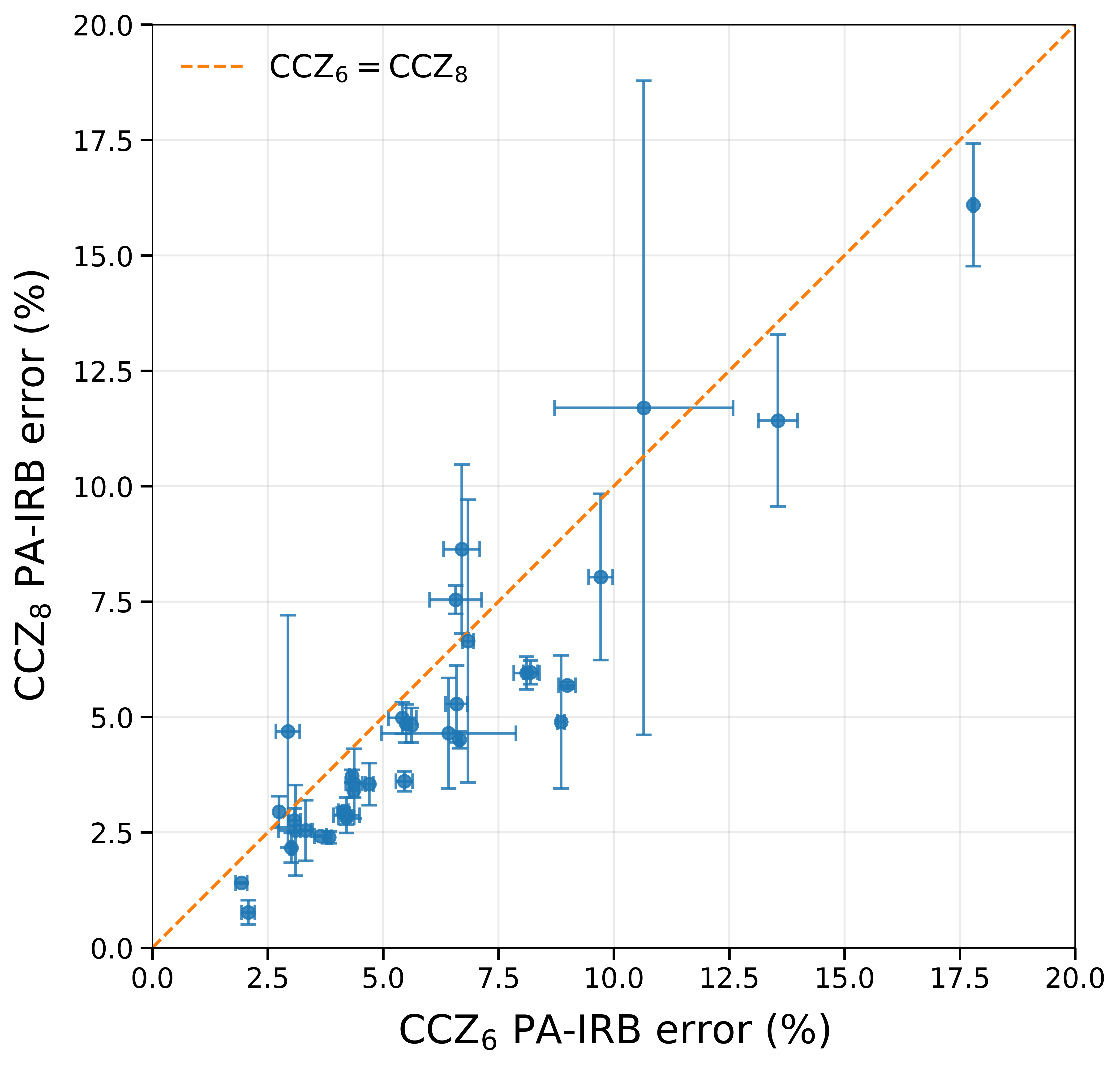}
\caption{
Phase-altered interleaved randomized benchmarking comparison of the
two compiled CCZ decompositions on \texttt{ibm\_kingston}. The
horizontal coordinate corresponds to the nominal $\cczs$ decomposition,
which compiles to 12 native CZ gates on the linear physical topology,
and the vertical coordinate corresponds to the $\cczl$ LNN
decomposition, which retains eight native CZ gates. Each marker
represents one physical three-qubit placement and is positioned at the
midpoint of the phase-stripped and phase-dressed PA-IRB estimates.
Horizontal and vertical bars span the corresponding stripped--dressed
values for the $\cczs$ and $\cczl$ realizations, respectively; these spans
are not statistical standard deviations. Only placements for which all
four PA-IRB estimates lie between 0 and 20\% are shown. The dashed line
denotes equal error; points below it favor the topology-compatible
$\cczl$ realization.
}
\label{fig:pairb}
\end{figure}

The reference and interleaved survival probabilities are fitted to
the usual exponential model
\begin{equation}
P(m)=A\alpha^m+B,
\label{eq:pairb_decay}
\end{equation}
where $\alpha$ is the fitted decay parameter. Denoting the reference
decay on physical triple $j$ by $\alpha_{\mathrm{RB},j}$ and the
interleaved decay by $\alpha_{k,v,j}$, with
$k\in\{6,8\}$ and $v\in\{s,d\}$, we define
\begin{equation}
r_{k,v,j}
=
\frac{d-1}{d}
\left(
1-\frac{\alpha_{k,v,j}}
        {\alpha_{\mathrm{RB},j}}
\right)=
\frac{7}{8}
\left(
1-\frac{\alpha_{k,v,j}}
        {\alpha_{\mathrm{RB},j}}
\right),
\label{eq:pairb_error}
\end{equation}
with $ d=2^3=8$. 
These quantities are PA-IRB estimates associated with the Clifford
surrogates of the two compiled decompositions. 
They are not direct
IRB estimates of the error of the non-Clifford CCZ unitary, and their
midpoints below are descriptive summaries rather than process-infidelity
estimators.

For each decomposition, the stripped and dressed estimates provide a
pair of operational values, $\{r_{6,s,j},r_{6,d,j}\}$ and $\{r_{8,s,j},r_{8,d,j}\}$.
For visualization we define their midpoints
\begin{equation}
\bar r_{6,j}
=
\frac{r_{6,s,j}+r_{6,d,j}}{2},
\quad
\bar r_{8,j}
=
\frac{r_{8,s,j}+r_{8,d,j}}{2},
\label{eq:pairb_midpoints}
\end{equation}
and the corresponding half-spans
\begin{equation}
w_{6,j}
=
\frac{|r_{6,d,j}-r_{6,s,j}|}{2},
\quad
w_{8,j}
=
\frac{|r_{8,d,j}-r_{8,s,j}|}{2}.
\label{eq:pairb_spans}
\end{equation}
The horizontal and vertical bars in Fig.~\ref{fig:pairb} represent
these stripped-dressed spans. They are therefore not statistical
one-standard-deviation error bars; rather, they show the variation of
the extracted IRB estimate under the two phase alterations of the same
compiled decomposition.

For the comparative scatter plot we retain placements for which all
four PA-IRB estimates lie in the interval
$0\le r_{k,v,j}\le 0.20$.
Placements outside this range are classified as outliers for the present comparison and are omitted from the displayed scatter plot, while their raw RB and IRB data and fitted parameters are retained in the experimental data set. 
This cut also removes cases in which an ill-conditioned decay fit produces a negative or otherwise non-informative IRB ratio.

Figure~\ref{fig:pairb} provides a decay-based comparison.
Because the two decompositions are benchmarked on the same physical triples and against the same reference RB procedure, their relative location with respect to the equality line compares the corresponding Clifford surrogates under repeated interleaving. 
Among the 27 retained placements, the mean midpoint error decreases from approximately $5.93\%$ for $\cczs$ to $4.94\%$ for $\cczl$, with 22 placements favoring $\cczl$. 

\section{Discussion}

Tables \ref{tab:mean-performance} and \ref{tab:best-performance} summarize the average and best performance of the $\cczs$ and $\cczl$ implementations of the CCZ gate in the three separate experiments conducted: EFS, hypergraph state fidelity and PA-IRB.
All three experiments reveal a reversal between logical and physical optimality. 
At the circuit level, $\cczs$ is optimal in CX count. 
At the hardware level, the missing endpoint coupling changes the executable resource hierarchy: the $\cczs$ template becomes the longer twelve-CZ implementation, whereas $\cczl$ remains an eight-CZ circuit. 
The experimentally measured EFS comparison demonstrates this reversal, and the independent hypergraph-state measurements confirm this conclusion on two processors. 
PA-IRB supplies a third, decay-based comparison through Clifford surrogates of the two compiled structures.

\begin{table}[tbph]
\caption{Mean performance extracted from the measured data. 
For EFS and the hypergraph-state benchmark, fidelity is one minus the displayed infidelity. The PA-IRB values are one minus the mean midpoint error of the Clifford surrogates. 
}
\label{tab:mean-performance}
\centering
\begin{tabular}{llcc}
\toprule
{Benchmark} & {Processor} & {$\cczs$} & {$\cczl$}  \\
\midrule
{EFS fidelity} & {\texttt{ibm\_fez}}      & {$87.8\%$} & {$91.8\%$}  \\
{EFS fidelity} & {\texttt{ibm\_kingston}} & {$90.2\%$} & {$93.4\%$}  \\
{Hypergraph fidelity} & {\texttt{ibm\_fez}}      & {$86.5\%$} & {$91.1\%$}  \\
{Hypergraph fidelity} & {\texttt{ibm\_kingston}} & {$89.8\%$} & {$92.3\%$} \\
{PA-IRB} & {\texttt{ibm\_kingston}} & {$94.1\%$} & {$95.1\%$}  \\
\bottomrule
\end{tabular}
\end{table}

\begin{table}[tbph]
\caption{Best displayed cases extracted from the measured data. 
Each row compares $\cczs$ and $\cczl$ on the same physical placement. 
The PA-IRB row reports best midpoint values of Clifford surrogates. 
}
\label{tab:best-performance}
\centering
\begin{tabular}{llll}
\toprule
{Benchmark} & {Processor} & {$\cczs$} & {$\cczl$} \\
\midrule
{EFS} & {\texttt{ibm\_fez}, best T8} & {$94.4\%$} & {$96.9\%$} \\
{EFS} & {\texttt{ibm\_kingston}, best T8} & {$96.8\%$} & {$98.1\%$} \\
{Hypergraph} & {\texttt{ibm\_fez}, best T8} & {$94.8\%$} & {$96.7\%$} \\
{Hypergraph} & {\texttt{ibm\_kingston}, best T8} & {$97.9\%$} & {$99.0\%$} \\
{Hypergraph} & {\texttt{ibm\_kingston}, best T6} & {$98.7\%$} & {$95.5\%$} \\
{PA-IRB} & {\texttt{ibm\_kingston}} & {$98.2\%$} & {$99.5\%$} \\
\bottomrule
\end{tabular}
\end{table}

The measured advantage is relevant beyond one gate template. 
Abstract two-qubit gate counts become physically meaningful only after the topology, native basis, and compilation constraints have been specified. 
The present experiment isolates a complementary limit: no relative-phase or approximate substitution is used, both competing templates are exact, and the comparison is paired on the same physical triples.
This conclusion is consistent with architecture-tuned and context-aware Toffoli compilation \cite{Duckering2021,QContext}, native-gate optimization \cite{Bowman2023}, and connectivity-aware exact synthesis \cite{CruzMurta2024}. 


\section{Conclusion}

In conclusion, we have compared the CNOT-optimal $\cczs$ and topology-compatible LNN $\cczl$ exact Toffoli/CCZ decompositions on linear triples of two IBM Quantum Heron processors. 
Under the fixed compilation used here, the former becomes a 12-CZ physical circuit while the latter remains an 8-CZ circuit. 
Experimentally measured EFS estimates favor the LNN realization across nearly all retained triples, and the paired phase-sensitive hypergraph-state measurements favor it on most triples on both processors. 
PA-IRB measurements of corresponding Clifford surrogates provide consistent complementary evidence on \texttt{ibm\_kingston}. 
Within the scope of the tested circuits and phase-sensitive input state, the results demonstrate that sparse connectivity can reverse the operational ranking of exact quantum-circuit decompositions and motivate compiler objectives based on the physical resources actually used.

\begin{acknowledgments}
This research is supported by the Bulgarian national plan for recovery and resilience, contract BG-RRP-2.004-0008-C01 (SUMMIT: Sofia University Marking Momentum for Innovation and Technological Transfer), project number 3.1.4, and by the QuantERA project FALCON.

The authors acknowledge access to IBM Quantum processors. The views expressed are those of the authors and do not reflect the official policy or position of IBM or the IBM Quantum team.
\end{acknowledgments}

\appendix
\section{Physical qubit triples}
\label{app:triples}

The physical triples used on \texttt{ibm\_fez} and
\texttt{ibm\_kingston} are listed in
Tables~\ref{tab:triples-fez}, \ref{tab:triples-kingston}, and
\ref{tab:triples-pairb-kingston}. The EFS, hypergraph-state, and
PA-IRB measurements were performed in separate hardware sessions, and
the selected sets therefore differ slightly between the three
measurements. Each ordered triple is given as
$(q_0,q_1,q_2)$.

\begin{table}[tbph]
\caption{Physical triples used on \texttt{ibm\_fez}. The table lists
the 40 triples used for the EFS measurements. For the hypergraph-state
measurements, $(12,13,14)$ was replaced by $(11,12,13)$, and
$(124,125,126)$ was added, giving 41 triples in total.}
\label{tab:triples-fez}
\centering
\footnotesize
\setlength{\tabcolsep}{4pt}
\renewcommand{\arraystretch}{1.12}
\begin{tabular}{ccc}
\toprule
\multicolumn{3}{c}{EFS triples \texttt{ibm\_fez}} \\
\midrule
$(0,1,2)$       & $(3,4,5)$       & $(8,7,17)$ \\
$(12,13,14)$    & $(24,23,16)$    & $(25,37,45)$ \\
$(32,31,30)$    & $(34,33,39)$    & $(35,19,15)$ \\
$(36,21,20)$    & $(38,29,28)$    & $(41,42,43)$ \\
$(48,47,46)$    & $(49,50,51)$    & $(53,54,55)$ \\
$(56,63,62)$    & $(57,67,68)$    & $(65,77,85)$ \\
$(73,79,93)$    & $(74,75,59)$    & $(78,69,70)$ \\
$(82,81,76)$    & $(88,87,86)$    & $(89,90,91)$ \\
$(97,107,106)$  & $(100,101,116)$ & $(104,105,117)$ \\
$(109,110,111)$ & $(112,113,114)$ & $(115,99,95)$ \\
$(120,121,122)$ & $(123,136,143)$ & $(128,129,130)$ \\
$(132,131,138)$ & $(134,133,119)$ & $(140,141,142)$ \\
$(144,145,146)$ & $(147,137,127)$ & $(149,150,151)$ \\
$(154,155,139)$ &                 &                 \\
\bottomrule
\end{tabular}
\end{table}

\begin{table}[tbph]
\caption{Physical triples used on \texttt{ibm\_kingston}. The table
lists the 39 triples used for the EFS measurements. For the
hypergraph-state measurements, $(76,61,60)$, $(96,83,82)$,
$(137,147,148)$, and $(138,131,132)$ were replaced by
$(82,81,76)$, $(96,83,84)$, and $(147,137,127)$, giving
38 triples in total.}
\label{tab:triples-kingston}
\centering
\footnotesize
\setlength{\tabcolsep}{4pt}
\renewcommand{\arraystretch}{1.12}
\begin{tabular}{ccc}
\toprule
\multicolumn{3}{c}{EFS triples \texttt{ibm\_kingston}} \\
\midrule
$(3,4,5)$       & $(8,7,17)$       & $(11,12,13)$ \\
$(18,31,30)$    & $(24,23,22)$     & $(25,37,45)$ \\
$(32,33,39)$    & $(35,19,15)$     & $(36,21,20)$ \\
$(38,29,28)$    & $(41,42,43)$     & $(48,47,46)$ \\
$(49,50,51)$    & $(53,54,55)$     & $(56,63,62)$ \\
$(57,67,68)$    & $(65,77,85)$     & $(72,71,58)$ \\
$(73,79,93)$    & $(74,75,59)$     & $(76,61,60)$ \\
$(78,69,70)$    & $(88,87,86)$     & $(89,90,91)$ \\
$(96,83,82)$    & $(97,107,106)$   & $(100,101,116)$ \\
$(104,105,117)$ & $(109,110,111)$  & $(115,99,95)$ \\
$(123,136,143)$ & $(124,125,126)$  & $(128,129,130)$ \\
$(134,133,119)$ & $(137,147,148)$  & $(138,131,132)$ \\
$(140,141,142)$ & $(149,150,151)$  & $(154,155,139)$ \\
\bottomrule
\end{tabular}
\end{table}

\begin{figure*}[tbph]
\centering
\includegraphics[
    width=0.98\textwidth
]{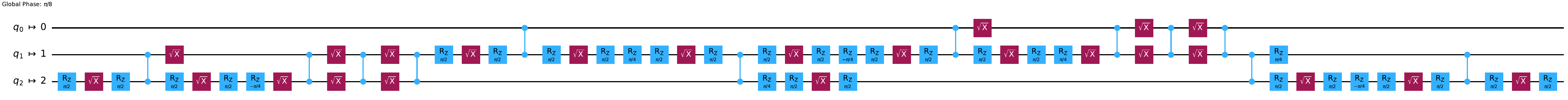}
\caption{
Native-basis transpilation of the nominal six-CX exact CCZ
decomposition on the linear topology $q_0-q_1-q_2$.
The non-nearest-neighbor interaction required by the logical
decomposition is routed through the central qubit, resulting in a
physical implementation containing 12 native CZ gates.
}
\label{fig:ccz6-native}
\end{figure*}

\begin{figure*}[tbph]
\centering
\includegraphics[
    width=0.98\textwidth
]{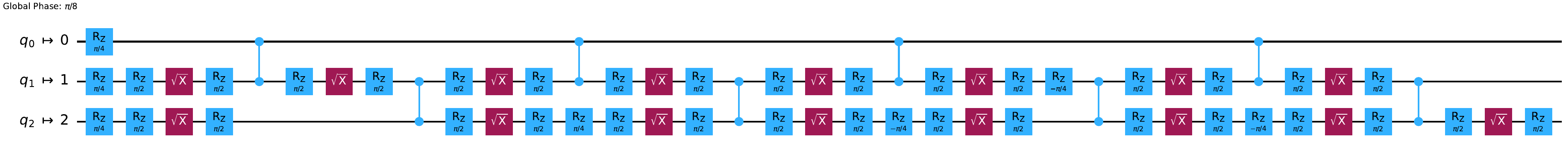}
\caption{
Native-basis transpilation of the eight-CX linear-nearest-neighbor
CCZ decomposition on the topology $q_0-q_1-q_2$.
Because all two-qubit interactions in the logical circuit already
respect the physical nearest-neighbor connectivity, the compiled
implementation contains 8 native CZ gates.
}
\label{fig:ccz8-native}
\end{figure*}

Figures~\ref{fig:ccz6-native} and \ref{fig:ccz8-native} show the
native-basis circuits obtained from the two exact CCZ decompositions
used in the experiments. The nominal six-CX and eight-CX circuits were
transpiled to the native unitary basis
$\{R_Z,SX,X,CZ\}$ on the linear three-qubit topology
$q_0-q_1-q_2$, using the same fixed compilation protocol throughout
the study. Because the six-CX decomposition contains an endpoint
interaction between $q_0$ and $q_2$, routing increases its physical
entangling-gate count to 12 native CZ gates. In contrast, the
eight-CX LNN decomposition uses only the two nearest-neighbor links
and retains eight native CZ gates after compilation. These compiled
circuits also provide the physical circuit structures from which the
phase-stripped and phase-dressed Clifford surrogates used in the
PA-IRB measurements were constructed.

\begin{table}[tbph]
\caption{Physical triples used for the phase-altered interleaved
randomized benchmarking (PA-IRB) measurements on
\texttt{ibm\_kingston}. The table lists the 35 triples retained in
the PA-IRB comparison shown in Fig.~\ref{fig:pairb}. Only placements
for which all four phase-stripped and phase-dressed PA-IRB error
estimates for the two CCZ decompositions lie in the interval
$0\leq r\leq 0.20$ are included.}
\label{tab:triples-pairb-kingston}
\centering
\footnotesize
\setlength{\tabcolsep}{4pt}
\renewcommand{\arraystretch}{1.12}
\begin{tabular}{ccc}
\toprule
\multicolumn{3}{c}{PA-IRB triples \texttt{ibm\_kingston}} \\
\midrule
$(0,1,2)$       & $(3,4,5)$        & $(8,7,17)$ \\
$(11,12,13)$    & $(18,31,30)$     & $(24,23,22)$ \\
$(25,37,45)$    & $(32,33,39)$     & $(35,19,15)$ \\
$(36,21,20)$    & $(38,29,28)$     & $(41,42,43)$ \\
$(48,47,46)$    & $(49,50,51)$     & $(53,54,55)$ \\
$(56,63,62)$    & $(57,67,68)$     & $(65,77,85)$ \\
$(72,71,58)$    & $(73,79,93)$     & $(74,75,59)$ \\
$(78,69,70)$    & $(82,81,76)$     & $(89,90,91)$ \\
$(96,83,84)$    & $(97,107,106)$   & $(100,101,116)$ \\
$(104,105,117)$ & $(109,110,111)$  & $(123,136,143)$ \\
$(124,125,126)$ & $(140,141,142)$  & $(147,137,127)$ \\
$(149,150,151)$ & $(154,155,139)$  & \\
\bottomrule
\end{tabular}
\end{table}


\bibliography{toffoli_references}
\end{document}